\RequirePackage{fix-cm}
\documentclass[twocolumn]{svjour3}          % twocolumn
\smartqed  % flush right qed marks, e.g. at end of proof
\usepackage{graphicx}
\usepackage{subfigure}
\usepackage{bbding}
\usepackage{array}
\usepackage{url}
\begin{document}

\title{Cross-Layer Software-Defined 5G Network
\thanks{This work is supported by National Basic Research Program of China (973 Program Grant No.~2013CB329105), National Natural Science Foundation of China (Grants No.~61301080, No.~61171065, 61271279, and 61201157),
Chinese National Major Scientific and Technological Specialized Project (No.~2013ZX03002001 and 2014AA01A707), China��s Next Generation Internet (No.~CNGI-12-03-007), and the Fundamental Research Foundation of NPU (Grant No. JCY20130131).}
}

\author{Mao Yang \and
        Yong Li \and
        Long Hu \and
        Bo Li \and
        Depeng Jin \and
        Sheng~Chen \and
        Zhongjiang Yan 
}


\institute{M. Yang $\cdot$ B. Li $\cdot$ Z. Yan \at
              School of Electronics and Information, Northwestern Polytechnical University, Xi'an 710072, P. R. China\\
              \email{yangmao@nwpu.edu.cn}
           \and
           Y. Li $\cdot$ D. Jin \at
              Department of Electronic Engineering, Tsinghua University, Beijing 100084, P. R. China\\
              \email{liyong07@tsinghua.edu.cn}
           \and
           L. Hu   \at
              School of Computer Science and Technology, Huazhong University of Science and Technology\\
              \email{minchen.cs@gmail.com}
           \and
           S. Chen \at
              Electronics and Computer Science, University of Southampton, Southampton SO17 1BJ, U.K.\\
              \email{sqc@ecs.soton.ac.uk}\\
}

\date{Received: date / Accepted: date}

\maketitle

\begin{abstract}
In the past few decades, the world has witnessed a rapid growth in mobile
 communication and reaped great benefits from it. Even though the fourth
 generation (4G) mobile communication system is just being deployed
 worldwide, proliferating mobile demands call for newer wireless
 communication technologies with even better performance. Consequently, the
 fifth generation (5G) system is already emerging in the research field.
 However, simply evolving the current mobile networks can hardly meet such
 great expectations, because over the years the infrastructures have
 generally become ossified, closed, and vertically constructed. Aiming to
 establish a new paradigm for 5G mobile networks, in this article, we
 propose a cross-layer software-defined 5G network architecture. By jointly
 considering both the network layer and the physical layer together, we
 establish the two software-defined programmable components, the control plane
 and the cloud computing pool, which enable an effective control of the
 mobile network from the global perspective and benefit technological
innovations. Specifically, by the cross-layer design for
software-defining, the logically centralized and programmable
control plane abstracts the control functions from the network layer
down to the physical layer, through which we achieve the
fine-grained controlling of mobile network,
 while the cloud computing pool provides powerful computing capability to
 implement the baseband data processing of multiple heterogeneous
 networks. The flexible programmability feature of our architecture makes
it convenient to deploy cross-layer technological innovations and
benefits the network
 evolution. We discuss the main challenges of our architecture, including
 the fine-grained control strategies, network virtualization, and
 programmability. The architecture significantly benefits the convergence
 towards heterogeneous networks and it enables much more controllable,
 programmable and evolvable mobile networks. Qualitative simulations
 validate these performance advantages.

\end{abstract}

\section{Introduction}\label{S1}

 With the rapid growth of mobile demands and the ever-increasing use of smart
 phones, mobile network has been one of the fastest growing technologies
 that impact major aspects of our life \cite{LTE_mobile_network_virtualization}.
 In recent years, the 4G mobile communication system is being deployed
 worldwide, leading to a rapid growth in the mobile network capacity, which
 further dramatically stimulates the mobile demands. As mobile network evolves
 from voice-oriented to media-oriented\cite{Multimedia_Centric_3G,Multimedia_Centric_Energy}, incremental improvements of networks
 can no longer keep the pace with the increase in mobile data demands, since
 there exist several challenges that the current mobile network can hardly address.

 \emph{1)~Convergence of heterogeneous wireless networks}. There exist various
 heterogeneous wireless networks, i.e. pluralistic standards such as GSM, UMTS,
 LTE and WLAN, and they will coexist for a long time \cite{Ericsson}. However,
 as current wireless operators typically deploy their networks by vertically
 constructed method, these heterogeneous networks can hardly interconnect with
 each other, which makes operators incapable of efficiently optimizing and
 dynamically scheduling from the global perspective. For example, although
 there may be many mobile networks around us, we may not be able to access
 the most appropriate one or select multiple networks simultaneously. The only
 thing we can do is to access one specific network all the time, even if this
 network performs quite poorly.

\emph{2)~Efficient utilization of wireless resources}. As a result of the
 difficulty in achieving the convergence of heterogeneous networks, many
 network devices are not fully utilized and plenty of wireless resources are
 wasted, while generating enormous amounts of energy consumption. While
 several big carriers, including AT\&T, Verizon, T-Mobile and Sprint, say that
 in the next few years they may not have enough spectrum to meet the demands
 for mobile data, many scientists and engineers, including Martin Cooper, father
 of the mobile phone, are convinced that the main reason for spectrum crisis
 is that the spectrum resources are not fully utilized \cite{Spectrum_Crisis}.

\emph{3)~Network innovations}. Proliferating mobile demands compel wireless
 operators to build sustained innovations to continuously boost their network
 performances. However, tightly coupling with specific hardware and lack of
 flexible control interfaces, the current mobile network can hardly provide a
 fast track for technological innovations.

\emph{4)~Wireless applications and services}. Wireless services proliferate
 significantly, e.g. Big Data \cite{BigData1,BigData2}, and different kinds of services require very different
 network characteristics. Unfortunately, the current mobile network often
 can only support these wide-ranging and very different services with the same
 network characteristics and, consequently, it may provide users with poor
 quality of service (QoS) and poor quality of experience (QoE).

 Because the current mobile network is vertically constructed over the years,
 it generally becomes ossified and closed, and has difficulty to meet the
 above mentioned challenges. This calls for the research and deployment of
 new generation of mobile network, specifically, 5G, which is capable of
 providing a more open and efficient mobile network. In particular, 5G should
 benefit from the convergence of heterogeneous networks, facilitate network
 evolution, boost the network performance, enhance users' QoS and QoE, and
 at the same time save resources and energy. These objectives require mobile
networks to offer much more efficient controlling and user data
processing functions across the network layer down to physical layer
by jointly considering
 the both layers together. However, since the physical layer technologies in
 mobile networks are much more complicated than in the wired scenarios,
implementing efficient control and programmable user planes across
the network layer
 and the physical layer is extremely challenging. The problem gets worse as
 the mobile environment becomes more complex, e.g. high mobility,
 frequent handovers, and heavy interferences. Consequently, the key
 technologies of 5G system are still open and attract increasingly attentions \cite{5G_Cache}.

 Software defined network (SDN), an innovative paradigm, advocates separating
 the control plane and data plane, and abstracting the control functions of
 the network into a logically centralized control plane, referred to as SDN
 controller \cite{SDN_Jennifer,CM_SDN_Management}. Aiming to establish a new
paradigm for 5G mobile network, in this article, we propose a
cross-layer software-defined 5G
 network architecture. In our architecture, the concept of
 SDN controller is extended to take the control functions of the physical
 layer into consideration as well, not just those of the network layer.
 Furthermore, we also introduce the novel cloud computing pool which considers
 the programmable user data processing functions of both the network layer and the physical
 layer jointly. Specifically, by abstracting the control functions of the both
 layers, we introduce a logically centralized programmable control plane oriented towards
 both network and physical layers, through which we achieve the fine-grained
controlling and flexible programmability of the both layers. For
example, we can control the packets
 forwarding in network layer as well as direct the beamforming and interference
 canceling in physical layer. The cloud computing pool proposed in our
 architecture provides powerful computing capability to implement the baseband
 data processing of multiple heterogeneous networks, which efficiently improves
 the convergence of heterogeneous wireless networks. Moreover, the programmable
 scheme in our architecture flexibly supports the network evolution and the
 deployment of any potential technological innovations which is particularly
 important as the key technologies of network and physical layers of
 5G. We also discuss the
 main technical challenges of our architecture, including the fine-grained
 control strategies oriented to the both network and physical layers, network
 virtualization that supports the customizability, and programmability that is
 beneficial to network evolution. Qualitative simulations are deployed to
 validate the advantages of our proposal. Our study demonstrates that our
 proposed 5G architecture is capable of achieving its design goals, namely,
 convergence of heterogeneous networks, fine-grained controllability, efficient
 programmability for network evolution, and network and service customizability,
 and therefore it offers an architecture design of future mobile network.

 The remainder of this article is organized as follows. We first offer a
 rethinking of mobile network, and this is followed by the introduction to
 SDN. Next, we present our cross-layer software-defined 5G architecture and
 analyze its features and system feasibility. Then we interpret the key
 technologies and challenges of our architecture. Several quantitative
 performance results are provided. Finally, we conclude the article by
 summarizing our contribution.

\section{Rethinking the Mobile Networking}\label{S2}

 The challenges to the current mobile network are mostly caused by the
 ossification, closeness, and vertically constructed features. Simply
 evolving the traditional networking architecture can hardly meet the
 requirements of future mobile networks. We rethink the problems of the
 current mobile networking and envisage how the future mobile network
 looks like, and consequently propose the design goals of our architecture.

\textbf{\emph{Ossified} \emph{vs} \emph{Controllable}}. Controlling in
 the current mobile network is limited and ossified. It is quite
 difficult to dynamically adjust the control strategies when the
 network status changes quickly. For example, many users with high
 mobility may go through numerous and frequent handovers, especially
 when the handovers occur among heterogeneous networks, which needs more
 fine-grained and dynamical controlling. Therefore, efficiently
 controlling is a fundamental function that the future network has to
 provide.

\textbf{\emph{Closed} \emph{vs} \emph{Programmable}}. Technological
 innovations in mobile network need to be continuously developed in order
 to meet the demands. Unfortunately, mobile network technologies are
 usually solidified in the hardware. For example, the unequal error
 protection (UEP), a physical layer technology, can provide higher video
 quality while consuming less spectrum. But to make such performance
 enhancing changes typically require replacing the
 hardware due to the tightly coupling with the hardware in the form of
 specific ASICs designed for each protocol of the LTE network
 \cite{OpenRadio}. In particular, dynamically deploying specific
 technologies based on the run-time network status becomes especially
 difficult. Therefore, it is necessary and highly desired to introduce
 greater programmability into mobile networks.

\textbf{\emph{Vertical} \emph{vs} \emph{Converging}}. The
 vertical-constructed method has dominated the mobile networking almost
 since the first generation mobile communication system was conceived,
 and it has brought great successes to mobile communication. However,
 as the scale of mobile network is quickly enlarging and the network
 becomes ever-increasingly dense, traditional vertical-constructed
 networking severely restricts the development of mobile networks. For
 example, 3G, LTE and WLAN can hardly interconnect, and the operators
 are incapable of controlling the network from the global perspective.
 Thus, enabling efficiently convergence of heterogeneous networks must
 be supported in the future mobile network.

\textbf{\emph{Unified} \emph{vs} \emph{Evolvable}}. Multiple mobile
 protocols and pluralistic standards will coexist for a long time
 \cite{Ericsson}. Each of them possesses the distinctive advantage for
 specific services or situations. For example, 3G provides better voice
 services and WLAN suits for video applications. The diversified world
 stimulates pluralistic demands, and thus it is almost impossible to find
 one unified network with solidified technologies to fullfill all the
 expectations. Consequently, the future mobile network must support smooth
 evolution for pluralistic standards.

 Based on the above analysis, we can envisage the basic characteristics of
 the future mobile network and naturally define the \emph{\textbf{design goals}}
 of our proposed 5G architecture, which are summarized as follows.

\begin{itemize}
\item Beneficial to the convergence of heterogeneous networks;
\item Fine-grained controllability for mobile networks;
\item High programmability and supply open interfaces for network technological
 innovations;
\item High evolvability and fast deployment;
\item Customizability of networks and services.
\end{itemize}

\section{Software Defined Network}\label{S3}

 SDN has become one of the hottest topics in computer networking. By separating
 the control plane and the data plane, SDN abstracts the control functions of
 the network layer into a logically centralized control plane, known as the SDN
 controller \cite{SDN_Jennifer,CM_SDN_Management}. This SDN controller makes the
 rules and control the behaviors of data plane devices from the global
 perspective. OpenFlow \cite{SDN_Openflow} is a most common
 realization of SDN. Many manufacturers, such as HP, NEC, IBM, produce the
 commercial OpenFlow switch, and several kinds OpenFlow controllers, e.g. NOX and
 floodlight, are available. The OpenFlow protocol is sustainably released by Open
 Networking Foundation (ONF), and SDN is becoming ever-increasingly popular in
 networking today.

\begin{table*}[!htbp]
%\vspace*{-6mm}
\caption{Comparison between the existing works and our architecture.}
\label{TAB1}
%\vspace*{-10mm}
\begin{center}
\begin{tabular}{r@{}lr@{}lr@{}lr@{}lr@{}l} \hline
\multicolumn{2}{c}{} &&network architecture&&physical layer controlling&&programmability
 &&scenario \\ \hline
\multicolumn{2}{c}{OpenRadio} && $\surd$ $\times$ just data plane && $\times$ not involve&& $\surd$
 && $\times$ not involve \\
\multicolumn{2}{c}{CellSDN}   && $\surd$                          && $\surd$             &&$\times$ not involve
 && $\times$ LTE oriented \\
\multicolumn{2}{c}{SoftRAN}   && $\surd$ $\times$ but macro-cell situation && $\surd$             && $\times$ not involve
 && $\times$ LTE oriented  \\
\multicolumn{2}{c}{OpenRF}   && $\times$                          && $\surd$             && $\times$ not involve
 && $\times$ WiFi oriented  \\
\multicolumn{2}{c}{SoftCell} && $\surd$                           && $\times$ not involve && $\surd$
 && $\times$ core network  \\
\multicolumn{2}{c}{MobiFlow}  && $\surd$                          && $\times$ not involve &&$\times$ not involve
 && $\times$ not involve \\
\multicolumn{2}{c}{Our Architecture}  &&$\surd$ &&$\surd$ &&$\surd$ &&heterogeneous network  \\ \hline
\end{tabular}
\end{center}
%\vspace*{-3mm}
\end{table*}

 In recent years, several researchers attempt to extend SDN into wireless
 and mobile networks. OpenRadio \cite{OpenRadio} mostly focuses on the
 programmable data plane design of software-defined wireless network, but
 this work does not propose the holistic network structure with the control
 plane architecture. Li \emph{et al.} \cite{CellSDN} present a
 software-defined cellular network architecture, CellSDN, that covers both
 access network and core network. However, this work is oriented to the LTE
 network, and it can hardly be adapted to the circumstances of heterogeneous
 networks. SoftRAN \cite{SoftRan} is a software-defined control plane of the
 radio access network by introducing a virtual big-base station, but this
 architecture mostly focuses on the one-tier situation, consisting of
 micro-cells, which is not suitable for the scenarios of heterogeneous
 networks. OpenRF \cite{OpenRF} and SoftCell \cite{SoftCell} focus on the software defined WiFi network and
 core network respectively. Meanwhile, MobiFlow \cite{MobiFlow} and OpenRoad \cite{OpenRoad} focus on
 the transport network that is similar to wired network and offer no special
 support for mobile access network. Moreover, almost all the existing works,
 except for OpenRadio and OpenRF, neglect the vital programmability features.
 Table~\ref{TAB1} compares our proposed architecture with these existing works.

\begin{figure}[htp]
%\vspace*{-8mm}
\begin{center}
 \includegraphics[width=0.4\textwidth]{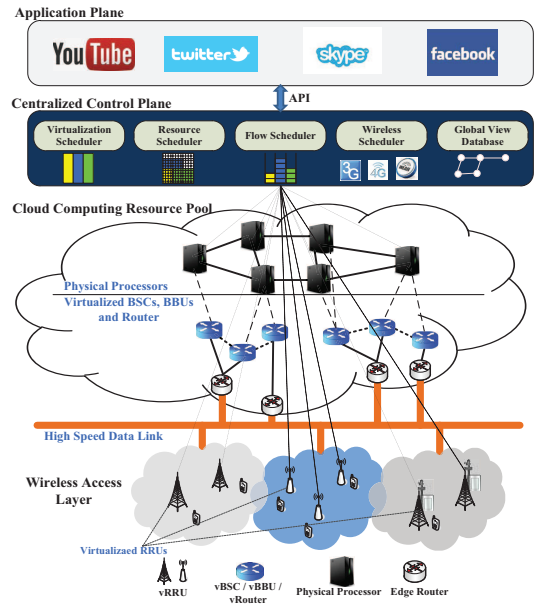}
\end{center}
%\vspace*{-10mm}
\caption{Overview of our proposed 5G architecture.}
\label{fig:architecture} % Fig 1
%\vspace*{-2mm}
\end{figure}

\section{Proposed 5G Architecture}\label{S4}

\subsection{Architecture Overview}\label{S4.1}

 In order to address the challenges and to achieve the goals introduced
 in the previous sections, we propose a cross-layer software-defined 5G
 architecture, as depicted in Fig.~\ref{fig:architecture}. We introduce
 a logically centralized control plane, which makes the rules and
 controls the behaviors of network devices as well as provides appropriate
 application program interfaces (APIs) for various applications. To achieve
 the efficient controlling of both the network layer and physical layer, we
 design a fine-grained control strategy which is compatible with the both
 layers. Because the limitation of poor computing capability in baseband
 data processing and the lack of interconnection severely impede the
 traditionally designed physical layer's controllability, we introduce a
 novel cloud based baseband data processing pool. The radio transceiver
 devices and the baseband data processing devices are completely separated.
 We also introduce the programmability feature to make the network more
 evolvable and flexible for technological innovations. Our architecture
 contains four main parts: wireless access layer, cloud computing resource
 pool, centralized control plane, and application plane.

 Wireless access layer consists of a large number of physical remote
 radio units (pRRUs) distributed at various locations. It enables several
 virtual RRUs (vRRUs) corresponding to different mobile communication
 protocols coexisting in one shared pRRU by radio frequency virtualization
 technology, to efficiently support fast deployment and the convergence of
 heterogeneous networks. For example, one pRRU can fast deploy one 3GPP
 vRRU and one WLAN vRRU (vAP) simultaneously, according to the requirements.

 Cloud computing resource pool is comprised of a large amount of physical
 processors with high-performance computing and storage capability as
 well as high speed links. The current mobile network adopts the
 vertically constructed networking, in which one RRU is tightly coupled
 with one specific base band unit (BBU). To loosen this inefficient
 coupling, in our architecture, the traditional BBUs, base station
 controllers (BSCs) and routers are replaced by virtual ones (vBBUs,
 vBSCs, and vRouters) which are implemented in the shared physical
 processors through virtualization. These shared physical processors can
 simultaneously run multiple vBBUs, vBSCs, and vRouters by allocating
 appropriate computing resources and storage resources. Specifically,
 each vBBU can deploy appropriate wireless protocol softwares to
 implement the physical layer programmability. The corresponding
 virtual elements (vBSCs, vBBUs, vRouters, and vRRUs) constitute a
 complete wireless network with specific protocols. Our cloud based
 data processing pool significantly improves the computing capacity
 of baseband processing and overcomes the interconnection difficulty of
 traditional mobile networks, which greatly benefits the convergence
 of heterogeneous networks. To guarantee the run-time properties, the
 wireless access pool is connected to the cloud computing resource pool
 through high speed optical links. The wireless access layer and the
 cloud computing resource pool comprise the data plane of our architecture.

 The centralized control plane is the ``brain'' of our architecture by
 abstracting and combining the control functions of both the network and
 physical layers. The control plane can acquire the configurations and
 real-time status of the both layers to make the decision from the global
 perspective. We partition the control plane into several functional
 submodules.

\noindent
 \indent \emph{1)~Flow scheduler}, which is related to the network layer, is
 similar to the wired SDN controller. It makes the polices and schedules
 the behaviors of each virtual forwarding element, i.e. guiding them to
 process the packets.

\noindent
 \indent \emph{2)~Wireless scheduler}, which is oriented to the physical layer, is
 the ``highlight'' in our architecture. It can deploy appropriate wireless
 protocol softwares into each vBBU. Moreover, it also dynamically schedules
 the physical layer actions, for example, beamforming, power control,
 interference cancelling, massive MIMO and carrier aggregation.

\noindent
\indent \emph{3)~Virtualization}, which is an indispensable element of SDN, makes
 it possible to customize networks and services. Specifically, virtualization
 scheduler manages the strategies of virtualization. We
 introduce three kinds of virtualization -- flow space virtualization, cloud
 level virtualization, and spectrum level virtualization -- that conspicuously
 improve the network performance from different angles.

\noindent
\indent \emph{4)~Resource scheduler}. Because there are various resources in our
 architecture, including spectrum, computing, and storage, it is necessary to
 efficiently allocate these resources to different virtual elements, which is
 implemented in resource scheduler.

\noindent
\indent \emph{5)~Global-view database}. All the above-mentioned submodules need to know
 the static configurations and dynamic status of the whole network and, therefore
 we introduce a database with global view to collect these information.

 The application plane contains a variety of network applications, each of which
 utilizes the APIs abstracted by the control plane to guarantee its QoS and QoE.

\subsection{Meeting the Design Goals}\label{S4.2}

 We now emphasize that our proposed cross-layer software-defined 5G architecture
 achieves the design goals set out in Section~\ref{S2}. Firstly, we abandon the
 traditional vertically constructed networking method and introduce a cloud
 computing pool to process the baseband data of heterogeneous networks. Therefore,
 our architecture conspicuously benefits the convergence of heterogeneous networks.
 With our architecture, the mobile network can easily acquire real-time global
 network view, which enables its centralized control plane to efficiently control
 the entire network by considering both the network and physical layers as a whole.
 This makes the network much more open and controllable. The flexible
 programmability proposed not only provides the opportunities for fast
 deployment of technological innovations but also makes the network more evolvable.
 Our proposed three types of virtualization further enhance the customizability of
 the networks and services.

\subsection{Feasibility Analysis}\label{S4.3}

 Recent years, SDN and network virtualization attract increasingly attentions from
 both the academic and industrial communities. Many  prototypes, testbeds and products
 have been developed and produced, in particular, OpenFlow controller, switch and
 protocols, as well as GENI network virtualization platform. Furthermore, data
 center network and cloud computing also stimulate fast development of SDN and
 virtualization technologies. Moreover, there are plenty of prototypes and testbeds
 for software defined radio (SDR) \cite{SDR} as well as its extended version,
 cognitive radio \cite{Cognitive_Radio}, which offer considerable programmability. We
 also note that China Mobile Ltd. positively focuses on the research and development
 of the cloud based mobile network architecture, referred to as CRAN \cite{C_RAN}.
 Based on these progresses, we believe that our proposed architecture is quite
 feasible.

\begin{figure}[htp]
%\vspace*{-7mm}
\begin{center}
 \includegraphics[width=0.4\textwidth]{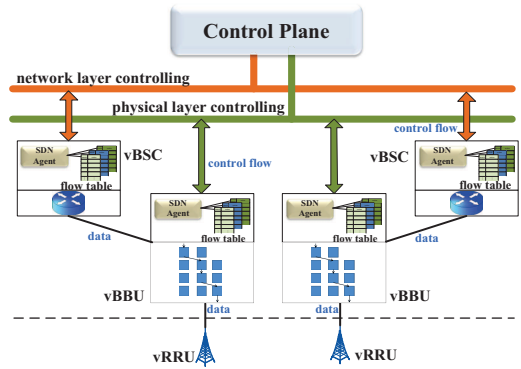}
\end{center}
%\vspace*{-10mm}
\caption{The proposed fine-grained control strategy architecture.}
\label{fig:control}  % Fig 2
%\vspace*{-3mm}
\end{figure}

\section{Key Challenges and Technologies}\label{S5}

\subsection{Fine-Grained Control Strategy}\label{S5.1}

 The first technical challenge is to design the fine-grained control strategy,
 and we divide the problem into two subproblems: jointly optimizing strategy
 and individual control strategies.

 The jointly optimizing strategy makes the decision to achieve the efficient
 controlling  according to the static configurations and real-time status of both
 the network and physical layers. Consider for example that the QoS for a user at
 cell boundary deteriorates. We may need to determine the appropriate routing and
 bandwidth requirements as well as to schedule the beamforming and interference
 cancelling. After the decision has been made, the control plane decomposes the
 decision into two parts corresponding to the two layers, and then leverages the
 flow based ``match-action'' strategy for the both layers to make rules and control
 behaviors of each virtual element, as illustrated in Fig.~\ref{fig:control}.
 Correspondingly, each data plane device integrates a SDN agent, which complies
 with the control strategy, resolves the control flow, and communicates with the
 control plane. Virtual elements constantly report the real-time status to the
 control plane, which may then dynamically refresh the global-view database.

 Since the responsibilities of the network layer and physical layer are
 distinctly different, the individual control strategies for the two layers are
 different, which defines the second subproblem of our fine-grained control
 strategy.

 \emph{Network Layer}. Since the packet header of each flow possesses several
 function fields called \emph{match fields}, such as IP address and MAC address,
 the data plane devices can process the arriving packets by checking
 the match fields. Specifically, when a virtual element receives a packet, it
 first checks whether this flow matches its control rules. If so, it executes
 the corresponding actions. Otherwise, the packets will be dropped or sent to
 the control plane. An example can be interpreted as: \textbf{if} dest
 IP = xx.xx.xx.xx, \textbf{then} forward to vBBU1 with speed S1.

 \emph{Physical Layer}. Although the physical layer controlling also adopts the
 ``match-action'' strategy, the match fields, rules, and actions are quite
 different from the network layer. Since key technologies of 5G physical layer
 are still open, the format of match fields should be elastic. For example,
 the basic information filed should contain the encoding mode, modulation type,
 carrier information, \emph{etc}. Additionally, the power control field may
 present the power control information, while the beamforming field and the
 interference cancelling field should provide some corresponding information. The
 control plane dynamically makes the decision according to the run-time
 status, and then sends control flows to the vBBUs. The ``actions'' in vBBUs
 refers to the appropriate physical layer technologies. For example, when QoS of
 a user deteriorates, the control plane may launch beamforming, while when the
 users at cell boundaries encounter severe inter-cell interference, the control
 plane may start interference cancelling.

\begin{figure}[htp]
%\vspace*{-6mm}
\begin{center}
\includegraphics[width=0.4\textwidth]{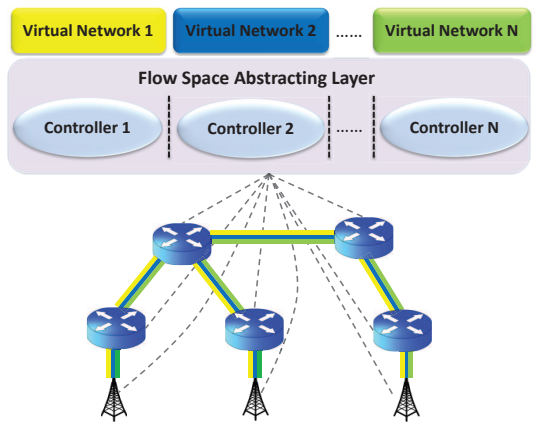}
\end{center}
%\vspace*{-10mm}
\caption{Flow space virtualization for networks or services.}
\label{fig:virtualization} % Fig 3
%\vspace*{-3mm}
\end{figure}

\subsection{Network Virtualization}\label{S5.2}

 Network virtualization as a promising means of the future network
 enables multiple concurrent virtual networks to run on the shared
 substrate resources \cite{Network_Virtualization,LTE_mobile_network_virtualization}.
 Network virtualization is an indispensable part of SDN, while SDN
 architecture facilitates network virtualization. Network virtualization
 is responsible for customizing the networks and services. Therefore, how
 the virtualization scheme is designed directly influences the system
 performance and resource utilization. There are three types of
 virtualization, flow space virtualization, cloud level virtualization,
 and spectrum level virtualization, in our architecture, and we now
 elaborate each of them.

 \emph{Flow space virtualization}. Different operators or network services
 require different network characteristics. As the control strategy in our
 architecture is flow based, flow space can be divided into multiple slices,
 each of which corresponds to one virtual network. As shown in
 Fig.~\ref{fig:virtualization}, we introduce a flow space abstracting layer.
 By abstracting the control functions of virtual elements, this layer slices
 the flow space according to the requirements of operators or services, and
 provides each flow slice with one controller. Then virtual networks are
 generated, and each virtual network is managed by its corresponding
 controller. The virtual networks share the same underlying data plane
 devices. In vBSC or vRouter, different virtual networks have different
 rules and run different actions. Operators often need different vBBUs due
 to the specific wireless protocols.

 \emph{Cloud level virtualization}. The control plane first creates vBBUs,
 vBSCs and vRouters by virtualizing physical processors and allocating
 appropriate computing and storage resources. Then it establishes the
 forwarding and data processing rules for virtual elements. Moreover, it
 deploys the corresponding wireless protocol softwares in vBBUs by utilizing
 the programmability.

 \emph{Spectrum level virtualization}. It refers to the virtualization of
 spectrum by radio frequency virtualization technology, which enables several
 vRRUs with different wireless protocols to coexist in one shared pRRU.
 Spectrum level virtualization extends the virtualization scope, lessens the
 networking difficulty and saves the cost.

\subsection{Programmability}\label{S5.3}

 The physical layer programmability faces several challenges including how to
 guarantee sufficient computing capability and how to efficiently implement
 physical layer technologies. Our proposed cloud based baseband processing
 pool provides powerful computing capability. In the physical layer, although
 different wireless protocols operate quite different from each other, they
 always share some same modules, such as modulation, coding and interleaving.
 Inspired from SDR, we modularize wireless protocols and establish a wireless
 module library in vBBUs. vBBUs can select different modules, and then connect
 these selected modules to implement the specific protocol. As 5G physical
 layer technologies are still open, our proposed programmability provides the
 flexibility for technological innovations, benefits the future network evolution,
 as well as enables fast deployment and customization of wireless protocols.

\section{Performance Evaluation}\label{S6}

 Although in the current stage the architecture proposed has not been deployed in a real scenario, we try to evaluate the performance with real traces from a large cellular deployment. We show that our proposed architecture significantly improves the energy efficiency and QoS utility.

\subsection{Dataset}\label{S6.1}
 To truly reflect the performance advantages, we use the real datasets named ``Sitefinder'' which is released by Ofcom on behalf of UK's government\cite{Ofcom}. These datasets record massive base stations information supplied by the mobile network operators. The information contains location information (latitude and longitude), operator information, base station type, transmission power, communication protocol (GSM, UMTS, and TETRA), frequency band, antenna information, and etc. Considering that the geographical area covered by these datasets is too wide, we select $10\emph{km} \times 10\emph{km}$ geographic area from Manchester. Consequently, there are 159 GSM base stations as well as 255 UMTS base stations deployed by five operators i.e. O2, Orange, Three, T-Mobile, and Vodafone.

\subsection{Energy Efficiency}\label{S6.2}
 In recent years, many researchers focus on the energy efficiency in mobile communications and, as a result, lots of energy-efficient mechanisms are presented. However, a form of stalemate exists: these efficient mechanisms are difficult to implement and deploy in real scenarios since the network functions are highly solidified in the hardware and the network devices are inflexible to control. Meanwhile, our proposed architecture provides flexible programmability and centralized controlling, which makes it easy to implement these energy-efficient schemes in real scenarios.

\begin{figure}[htp]
%\vspace*{-1mm}
\begin{center}
\subfigure[Energy Ratio for GSM]{\includegraphics[width=0.22\textwidth]{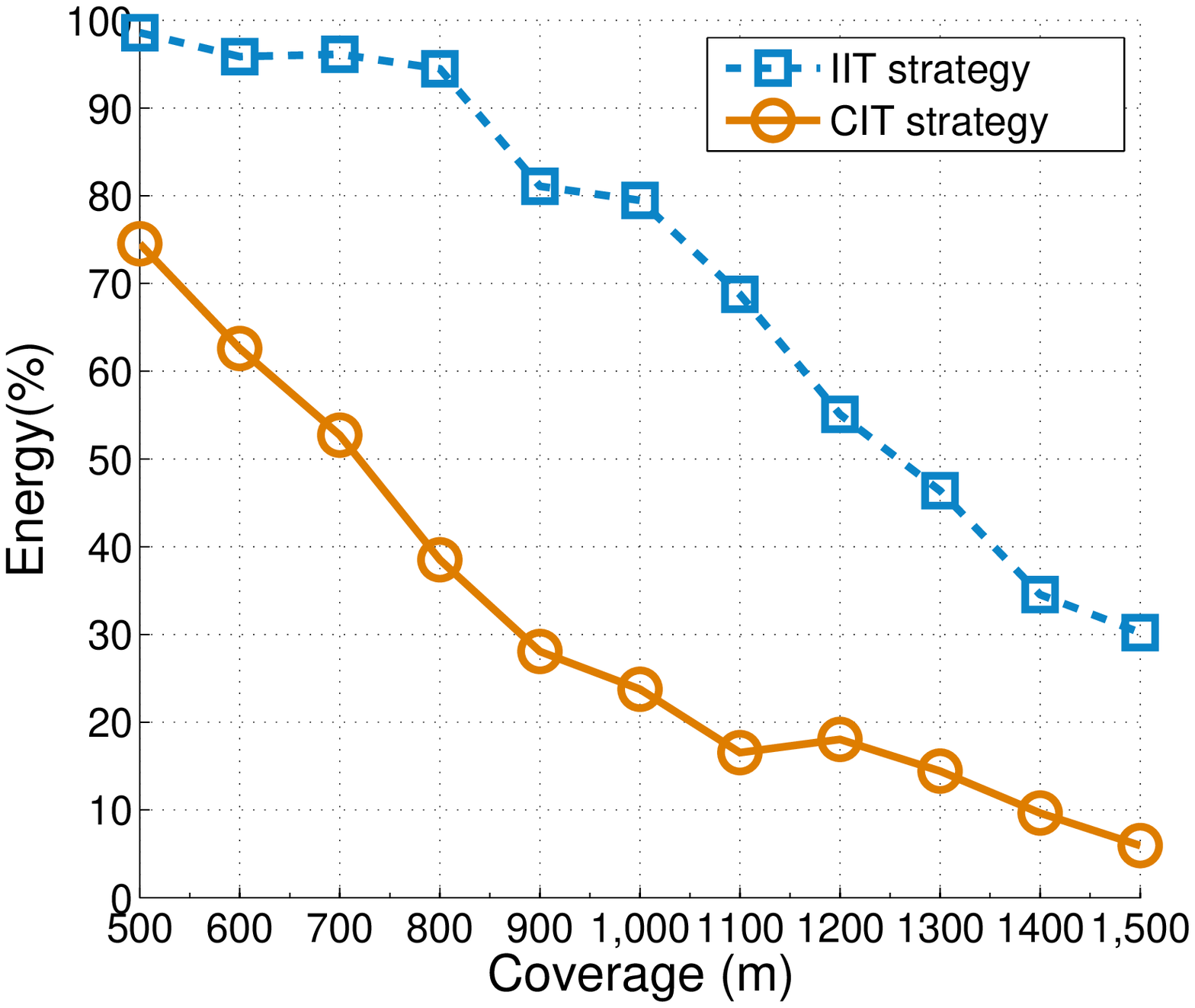}}
\subfigure[Base station Number for GSM]{\includegraphics[width=0.22\textwidth]{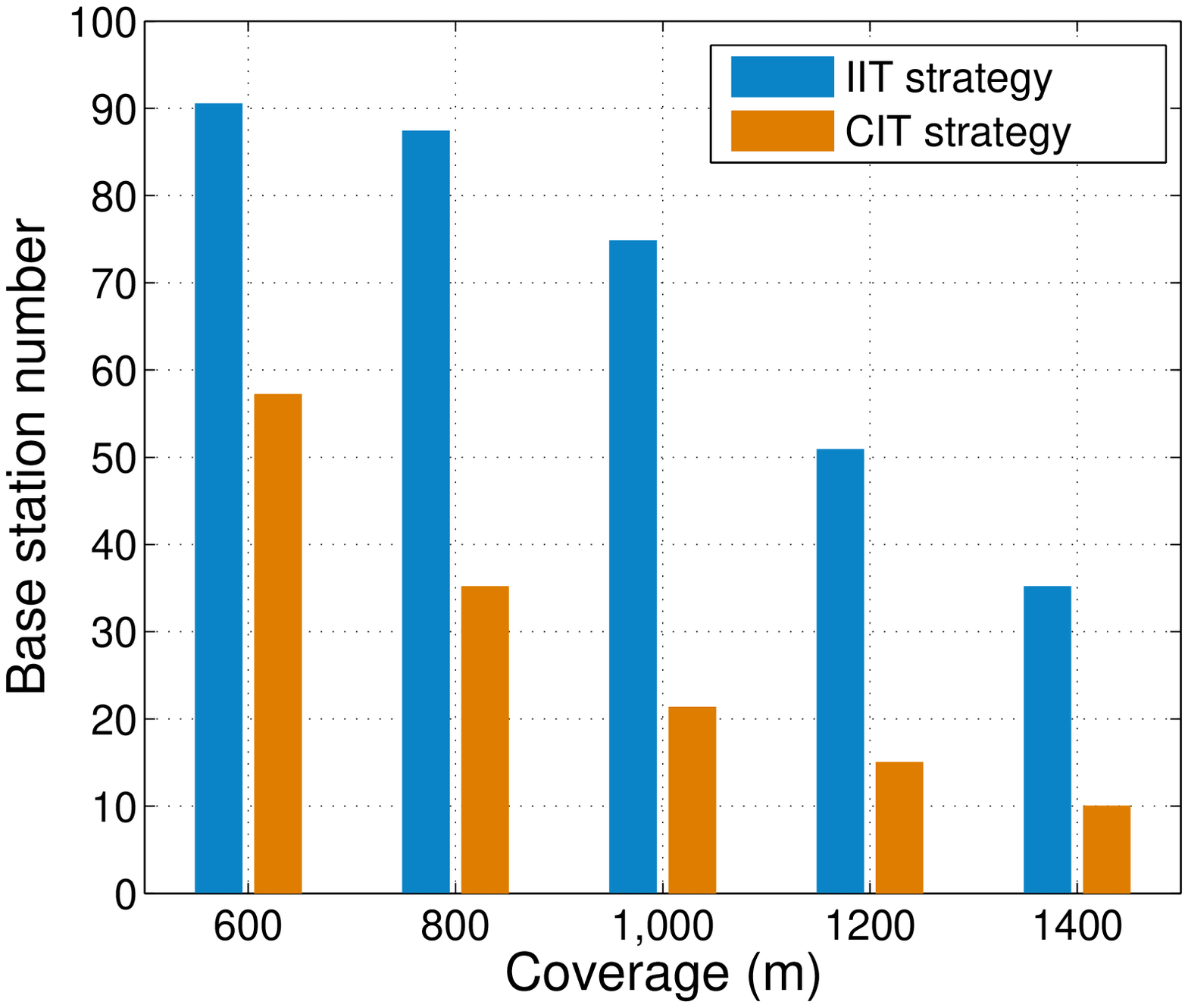}}
\subfigure[Energy Ratio for UMTS]{\includegraphics[width=0.22\textwidth]{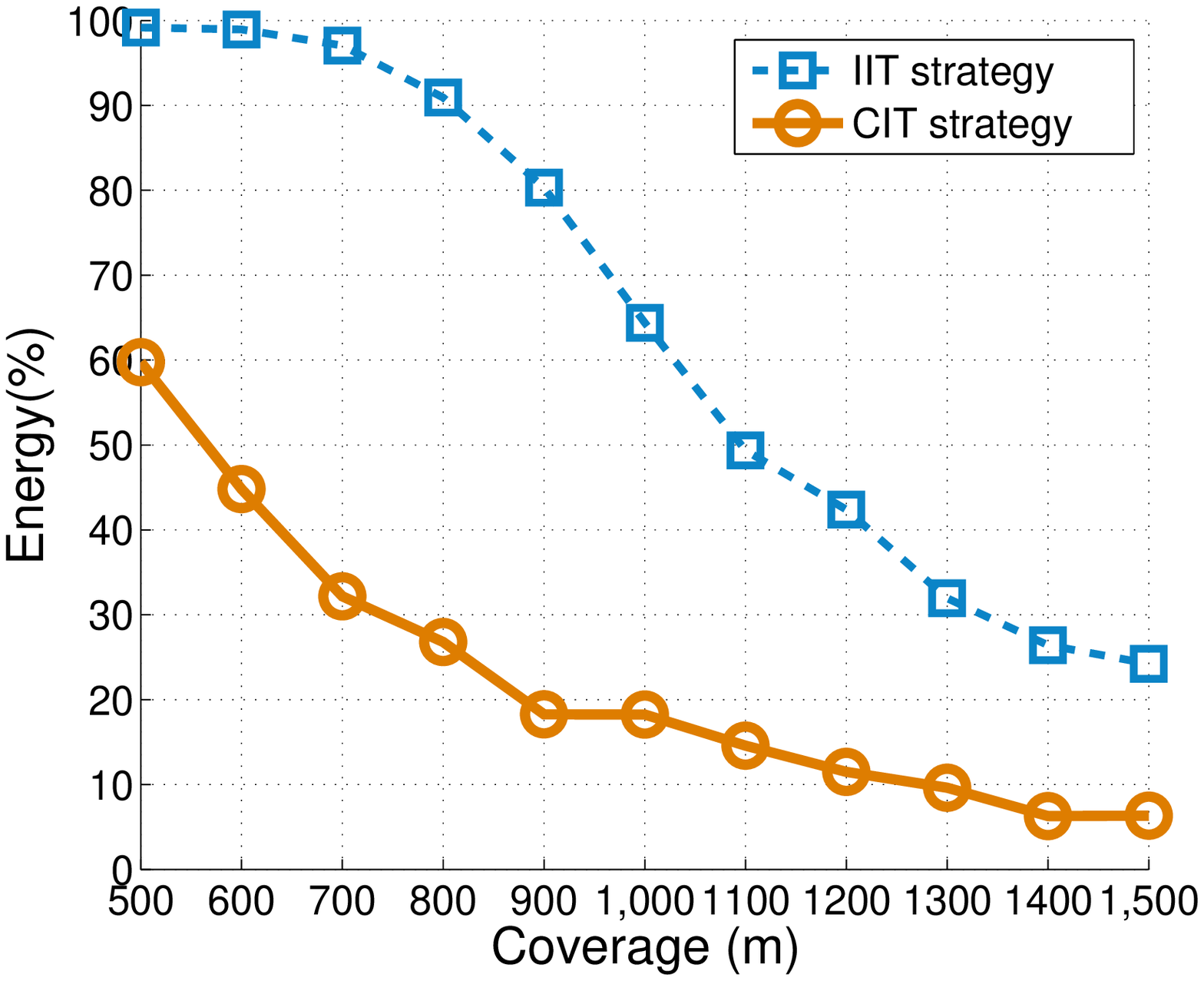}}
\subfigure[Base station Number for UMTS]{\includegraphics[width=0.22\textwidth]{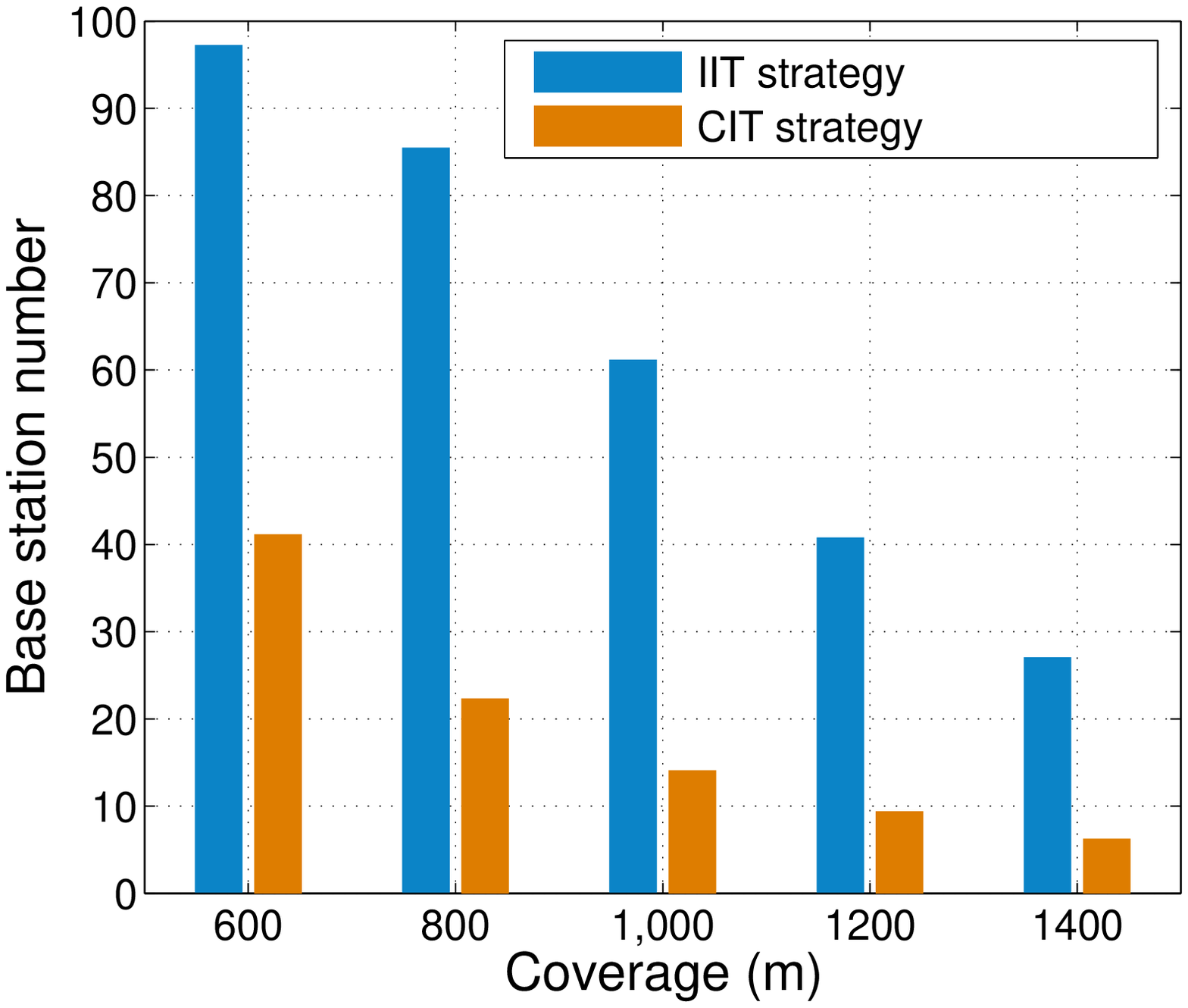}}
\end{center}
%\vspace*{-6mm}
\caption{Energy efficiency comparison.}
\label{fig:energy}  % Fig 4
%\vspace*{-7mm}
\end{figure}

 Considering that network traffic at midnight is low, i.e. tide effect, we propose two energy-efficient strategies: 1) the centralized control plane intelligently turn off the base stations on behalf of individual operator, referred to as IIT strategy; 2) control plane turn off the base stations from global perspective, i.e. cooperation cross operators, referred to as CIT strategy. It is notable that both of these two strategies needs to guarantee the whole area coverage. Fig.~\ref{fig:energy} depict the energy-saving performance of GSM and UMTS. As coverage range of each base station increases, both strategies turn more base stations and save more energy. For example, when coverage range is $1000\emph{m}$, 39 percent base stations can be turned off and 35 percent energy can be save by utilizing IIT strategy for UMTS. Moreover, as figures shown, apparently, the CIT strategy achieves much more energy-efficient benefits than IIT since CIT allows the control plane to optimize the energy-efficient solution across multiple mobile operators.

\subsection{QoS utility}\label{S6.3}

\begin{figure}[htp]
%\vspace*{-1mm}
\begin{center}
\subfigure[Strategy 1-3]{\includegraphics[width=0.22\textwidth]{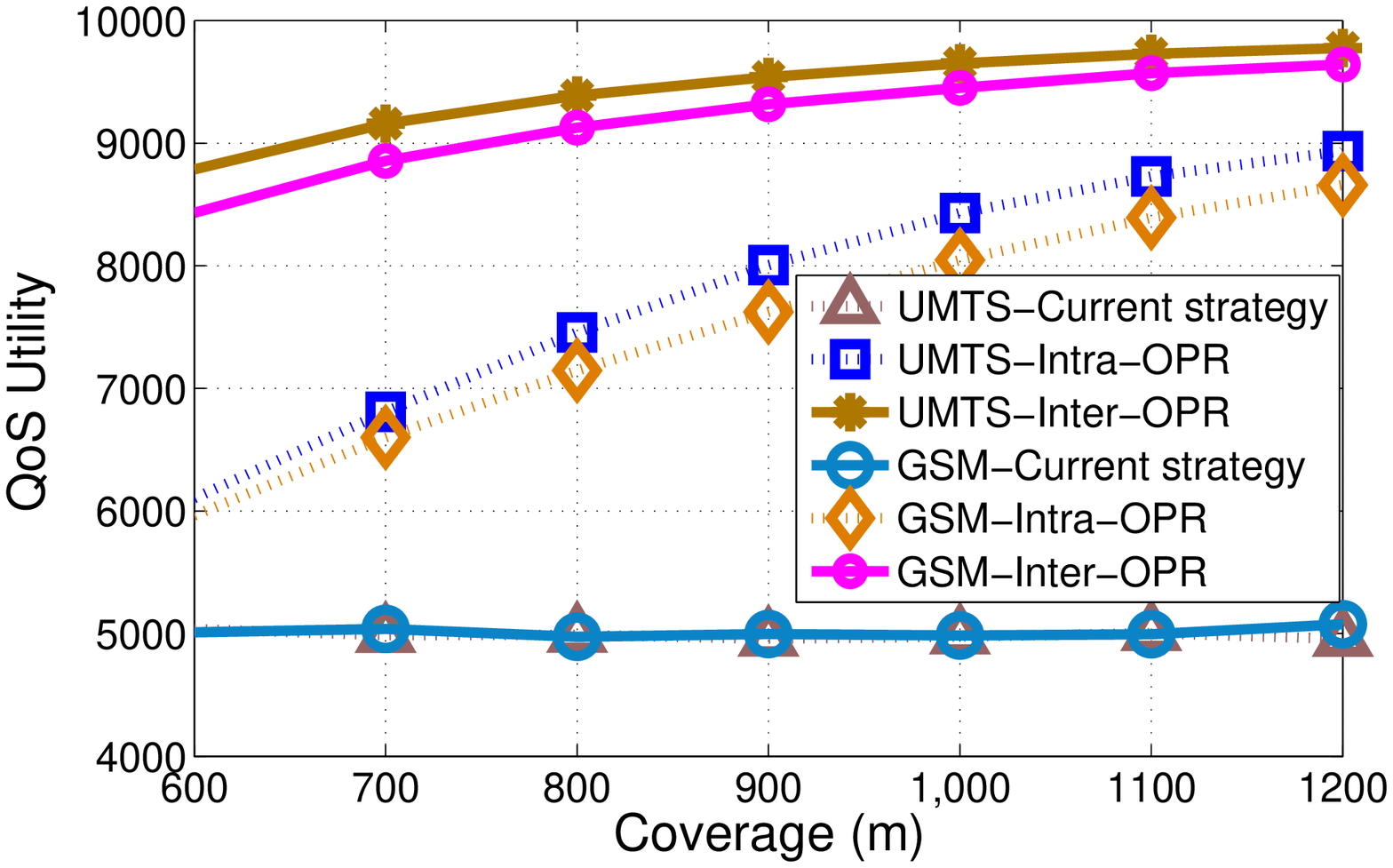}}
\subfigure[Strategy 4]{\includegraphics[width=0.22\textwidth]{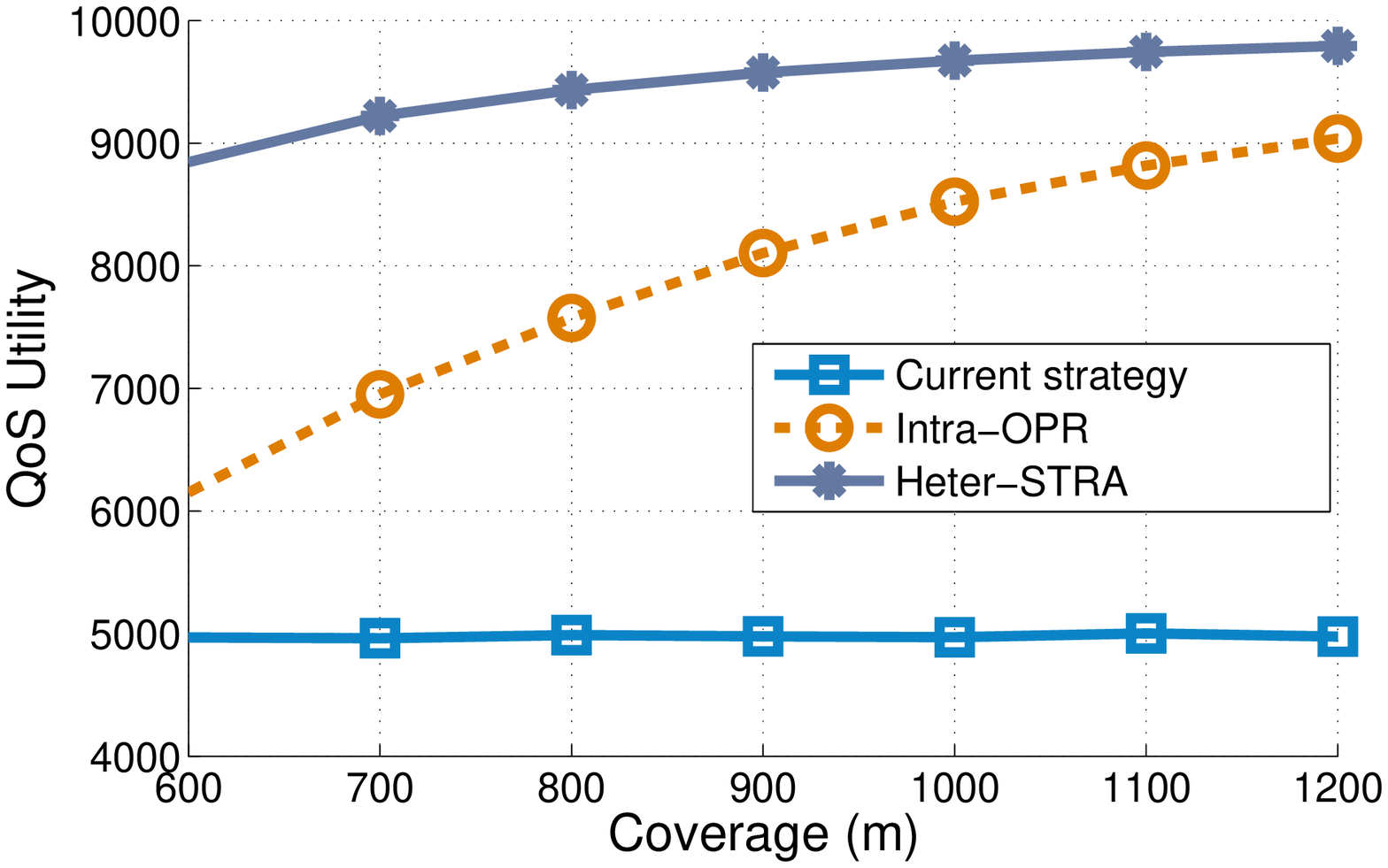}}
\end{center}
%\vspace*{-6mm}
\caption{QoS utility comparison.}
\label{fig:qos}       % Fig 5
%\vspace*{-7mm}
\end{figure}

 In the current mobile network, every end user has no choice but to access
 one specific network. They may not be able to access the most appropriate
 one or select multiple networks simultaneously, which leads to poor QoS.
 In contrast, in our architecture, all the network devices are programmable
 and controlled by the centralized control plane. Consequently, the control
 plane may guide the end users to access the most appropriate networks, by
 which QoS can be improved.

 We assume that there are 10,000 end users and each one runs an ongoing
 mobile service. We define the QoS utility between each pair of $<$end user,
 base station$>$, and assume the utility follows the uniform distribution in
 the range of $[0, ~1]$ by normalizing. The location (latitude and longitude)
 of each end user is randomly generated. We introduce four types of
 strategies: 1) Current strategy. Each business is bound to one specific
 network. 2) Intra operator optimizing (Intra-OPR): the control plane
 guides each end user to access the base station, belonging to the
 specific operator, with the largest QoS utility. 3) Inter operator
 optimizing (Inter-OPR): breaking the limit of one specific operator.
 4) Breaking the limit of operator as well as one communication standard
 (Heter-STRA), i.e. it allows the end users to the access the base station
 among several heterogeneous networks (in this senario: cross UMTS and GSM).
 As Fig.~\ref{fig:qos}(a) shows, both Intra-OPR and Inter-OPR obtain
 much more QoS utility than current strategy, and this advantage becomes
 much more obvious as the coverage range increases. After that, we consider
 the heterogeneous scenario and combine $159+255=414$ base stations.
 Fig.~\ref{fig:qos}(b) confirms the perform advantages of Heter-STRA.

\section{Summary and Conclusions}\label{S7}

 In this article, we have proposed a cross-layer software-defined 5G architecture,
 which addresses some of the key technical challenges facing the future mobile
 network. Our architecture is able to control the mobile network from the global
 perspective through the fine-grained controlling for the both network
 and physical layers. Moreover, the cloud computing pool in our architecture
 provides powerful computing capability to implement the baseband data
 processing of multiple heterogeneous networks, while the programmability
 feature efficiently supports the network evolution and technological innovations.

 The key technologies we introduced to our architecture include fine-grained
 control strategy, network virtualization, and programmability, and the key
 features of our proposed 5G architecture are controllability, programmability,
 evolvability and customizability. Since the key technologies of 5G are still
 open, we believe our flexible architecture is capable of supporting the 5G
 standards that eventually emerge and efficiently satisfying the requirements
 of mobile demands.

%\begin{acknowledgements}
%\end{acknowledgements}


\end{document}